%% file: main.tex
\documentclass[conference]{IEEEtran}
\usepackage{authblk}
\IEEEoverridecommandlockouts
\usepackage{cite}
\usepackage{amsmath,amssymb,amsfonts}
\usepackage{graphicx}
\input{preamble/math}

\usepackage{enumitem}
\usepackage[table]{xcolor}
\usepackage{multirow}
\definecolor{Gray}{gray}{0.9} 
\definecolor{LightBlue}{RGB}{173, 216, 230} 
\usepackage{textcomp}
\usepackage{xcolor}
\usepackage{graphicx} 
\usepackage{booktabs} 
\usepackage{amsmath}
\usepackage{amssymb}
\usepackage{algorithm}
\usepackage{algorithmic}
\usepackage[dvipsnames]{xcolor}
\usepackage{subcaption}
\usepackage{url}

\usepackage{array} 
\usepackage[most]{tcolorbox}
\usepackage[table,xcdraw]{xcolor} 
\usepackage{caption} 
\def\BibTeX{{\rm B\kern-.05em{\sc i\kern-.025em b}\kern-.08em
    T\kern-.1667em\lower.7ex\hbox{E}\kern-.125emX}}

\begin{document}

\date{16 Sept 2025}

\title{LIGHT-HIDS: A Lightweight and Effective Machine Learning-Based Framework for Robust Host Intrusion Detection}

\author{Onat Gungor$^{*}$}
\author{Ishaan Kale$^{*}$}
\author{Jiasheng Zhou}
\author{Tajana Rosing}

\affil{Department of Computer Science and Engineering, University of California, San Diego}
\affil{\{ogungor, ikale, rjzhou, tajana\}@ucsd.edu}

\maketitle

\begingroup
\renewcommand\thefootnote{\textasteriskcentered}
\footnotetext{Both authors contributed equally to this research.}
\endgroup

\pagestyle{plain}
\pagenumbering{gobble}
\newcommand{\norm}[1]{\left\lVert#1\right\rVert}
\newcommand{\Design}[0]{LIGHT-HIDS}

\begin{abstract}
The expansion of edge computing has increased the attack surface, creating an urgent need for robust, real-time machine learning (ML)-based host intrusion detection systems (HIDS) that balance accuracy and efficiency. In such settings, inference latency poses a critical security risk, as delays may provide exploitable opportunities for attackers. However, many state-of-the-art ML-based HIDS solutions rely on computationally intensive architectures with high inference costs, limiting their practical deployment. This paper proposes LIGHT-HIDS, a lightweight machine learning framework that combines a compressed neural network feature extractor trained via Deep Support Vector Data Description (DeepSVDD) with an efficient novelty detection model. This hybrid approach enables the learning of compact, meaningful representations of normal system call behavior for accurate anomaly detection. Experimental results on multiple datasets demonstrate that LIGHT-HIDS consistently enhances detection accuracy while reducing inference time by up to 75$\times$ compared to state-of-the-art methods. These findings highlight its effectiveness and scalability as a machine learning-based solution for real-time host intrusion detection.
\end{abstract}

\section{Introduction}
\input{sections/Intro.tex}

\section{Related Work}
\input{sections/RelatedWork.tex}

\section{\Design{} Framework}
\label{framework}
\input{sections/Framework.tex}

\section{Experimental Analysis}
\label{experimental}
\input{sections/Experiment.tex}

\section{Conclusion}
\label{conclusion}
\input{sections/Conclusion.tex}

\section*{Acknowledgments}
This work has been funded in part by NSF, with award numbers \#1826967, \#1911095, \#2003279, \#2052809, \#2100237, \#2112167, \#2112665, and in part by PRISM and CoCoSys, centers in JUMP 2.0, an SRC program sponsored by DARPA.

\bibliographystyle{ieeetr}
\bibliography{bibfile}

\end{document}

%% file: preamble/math.tex
\usepackage{xcolor}

%% file: sections/Intro.tex
The increasing reliance on digital infrastructure has elevated cybersecurity to a critical priority across diverse sectors. In response to the rising complexity and frequency of cyber threats, machine learning has become an essential tool for developing intelligent and adaptive defense mechanisms \cite{Kabbani2025}. A key application of machine learning within this domain is Host Intrusion Detection Systems (HIDS), where ML models are trained to analyze system-level data to identify potentially malicious activities \cite{Singh2025}. Among available data sources, system call traces have proven particularly effective due to their fine-grained insight into program execution behavior \cite{Banach2019}. Recent research has applied deep learning models, particularly those developed for natural language processing, to capture the sequential dependencies present in system call sequences \cite{Ring2021}. Although these models offer strong detection performance, their high computational complexity and inference latency hinder practical deployment in real-time scenarios.

These challenges become even more pronounced in resource-constrained environments such as IoT gateways, embedded controllers, and other edge computing platforms. Devices in these settings often operate with limited memory, compute power, and energy availability, making the deployment of conventional deep learning–based HIDS difficult without compromising responsiveness \cite{joshi2023enabling}. Moreover, edge devices frequently serve as the initial point of contact within a network, necessitating rapid, localized threat detection to reduce dependence on cloud infrastructure \cite{zhukabayeva2025cybersecurity}. Due to the strict real-time requirements of edge platforms, high inference latency in host intrusion detection systems can result in significant delays when analyzing system calls prior to program execution, thereby hindering timely operation \cite{Chen2025}. Additionally, during continuous monitoring, prolonged detection times increase the opportunity for attackers to exploit system vulnerabilities before anomalies are identified and addressed \cite{Singh2025}. These considerations underscore the importance of lightweight machine learning approaches that balance inference speed, model compactness, and detection accuracy to meet the latency and efficiency demands of practical edge deployments. While TinyML techniques have shown potential for enabling efficient intrusion detection on constrained devices~\cite{im2024tinyml, saranya2024secure, fusco2025tinyml}, existing research predominantly targets network telemetry rather than system call–based detection, highlighting a significant gap in securing edge-hosted systems.

To address these challenges, we propose LIGHT-HIDS (Figure~\ref{fig:framework}), a lightweight host-based intrusion detection framework. Our method employs a hybrid anomaly detection pipeline where system call sequences are first processed by a compressed neural network feature extractor trained via Deep Support Vector Data Description. The resulting compact and discriminative feature embeddings are then provided to an efficient novelty detector that assigns anomaly scores used for classification. Experimental evaluation shows that LIGHT-HIDS reliably improves detection accuracy while achieving inference time reductions of up to 75$\times$ relative to current state-of-the-art approaches. Unlike existing methods that are either too computationally intensive for practical edge deployment or too simplistic to effectively detect complex anomalies, LIGHT-HIDS achieves a balanced trade-off between efficiency and performance. To the best of our knowledge, this work represents one of the first lightweight and accurate HIDS designed with resource-constrained computing environments in mind, thereby addressing an important gap in edge security.

%% file: sections/RelatedWork.tex
\subsection{Host-based intrusion detection systems (HIDS)}
Host-based intrusion detection systems (HIDS) are a fundamental component of modern cybersecurity architectures, designed to monitor internal system activity and detect signs of malicious behavior~\cite{satilmics2024systematic}. Among the various types of host-level data, system calls are especially valuable, as they mediate interactions between user-space applications and the operating system kernel~\cite{liu2018host}. This fine-grained visibility makes system calls well-suited for capturing the underlying behavioral patterns of executing programs. While traditional signature-based HIDS effectively detect known threats, they struggle to identify new attacks. Anomaly-based approaches address this limitation by leveraging machine learning models to characterize normal system behavior and detect deviations indicative of malicious activity~\cite{martins2022host}. Such techniques are particularly effective in identifying zero-day exploits and evolving threats. 

Traditional anomaly-based HIDS methods, e.g., Isolation Forest~\cite{liu2008isolation}, depend heavily on manual feature engineering and face difficulties capturing complex temporal dependencies in system call sequences~\cite{buczak2015survey}. Deep learning addresses these limitations by automatically learning hierarchical features and effectively modeling intricate sequential patterns. Recurrent architectures like Long Short-Term Memory (LSTM) networks have been foundational in this area, excelling at capturing long-range dependencies and detecting anomalies from system calls~\cite{Kim2016}. Expanding on this foundation, hybrid models that integrate Convolutional Neural Networks (CNNs) with RNNs enhance performance by leveraging CNNs to extract local patterns and RNNs to capture temporal context~\cite{Chawla2018}. CNNs using dilated convolutions, such as those inspired by WaveNet, offer a parallelizable alternative that efficiently captures dependencies across varying temporal scales~\cite{Ring2021}. Transformer-based models have recently gained prominence in HIDS due to their self-attention mechanism, which simultaneously captures long-range relationships across entire sequences~\cite{Kunwar2025}. 

While Deep Learning (DL) methods offer powerful capabilities for modeling complex sequential patterns, they often come with substantial computational overhead that limits their suitability for real-time detection in resource-constrained environments. These models typically require large memory footprints and incur high inference latency due to their depth and complexity. Additionally, the sequential nature of system call data can lead to long input sequences, further increasing computation time and power consumption. Such constraints make it challenging to deploy conventional DL approaches directly on edge devices, highlighting the need for lightweight alternatives that can balance detection accuracy with efficiency.


\subsection{Hybrid Anomaly Detection}

Hybrid anomaly detection, which combines DL–based feature extraction with classical ML models, has been extensively studied~\cite{Chalapathy2019}. In this paradigm, DNNs transform raw input into compact, discriminative representations, reducing the need for manual feature engineering. These features are then leveraged by traditional algorithms, such as Isolation Forest or One-Class SVM, for anomaly detection. For example, autoencoders trained on normal data use reconstruction error or latent embeddings as input to classical detectors~\cite{Liu2021}. Another common approach employs CNNs followed by pooling layers to automatically extract hierarchical local patterns and distill salient features into compact representations suitable for anomaly detection~\cite{AlQatf2024}. We adopt this CNN-to-pooling strategy due to its demonstrated superior predictive performance.

Despite their success in general anomaly detection, these hybrid methods remain underexplored and insufficiently tailored to the unique challenges of host intrusion detection, such as the need for computational efficiency on resource-constrained devices, highlighting a critical gap that our work addresses. Building on this, we utilize DeepSVDD~\cite{Ruff2018}, which trains neural networks to map normal data into a compact hypersphere in latent space, concentrating normal samples near a learned center while pushing anomalies away. This approach facilitates learning compact representations of normal behavior that are highly effective for subsequent anomaly detection. These extracted features are then used to train a novelty detector, constituting a hybrid framework. To the best of our knowledge, we present the first efficient and effective hybrid solution specifically designed for HIDS.

%% file: sections/Framework.tex
\begin{figure*}[]
    \centering
    \includegraphics[width=.9\textwidth]{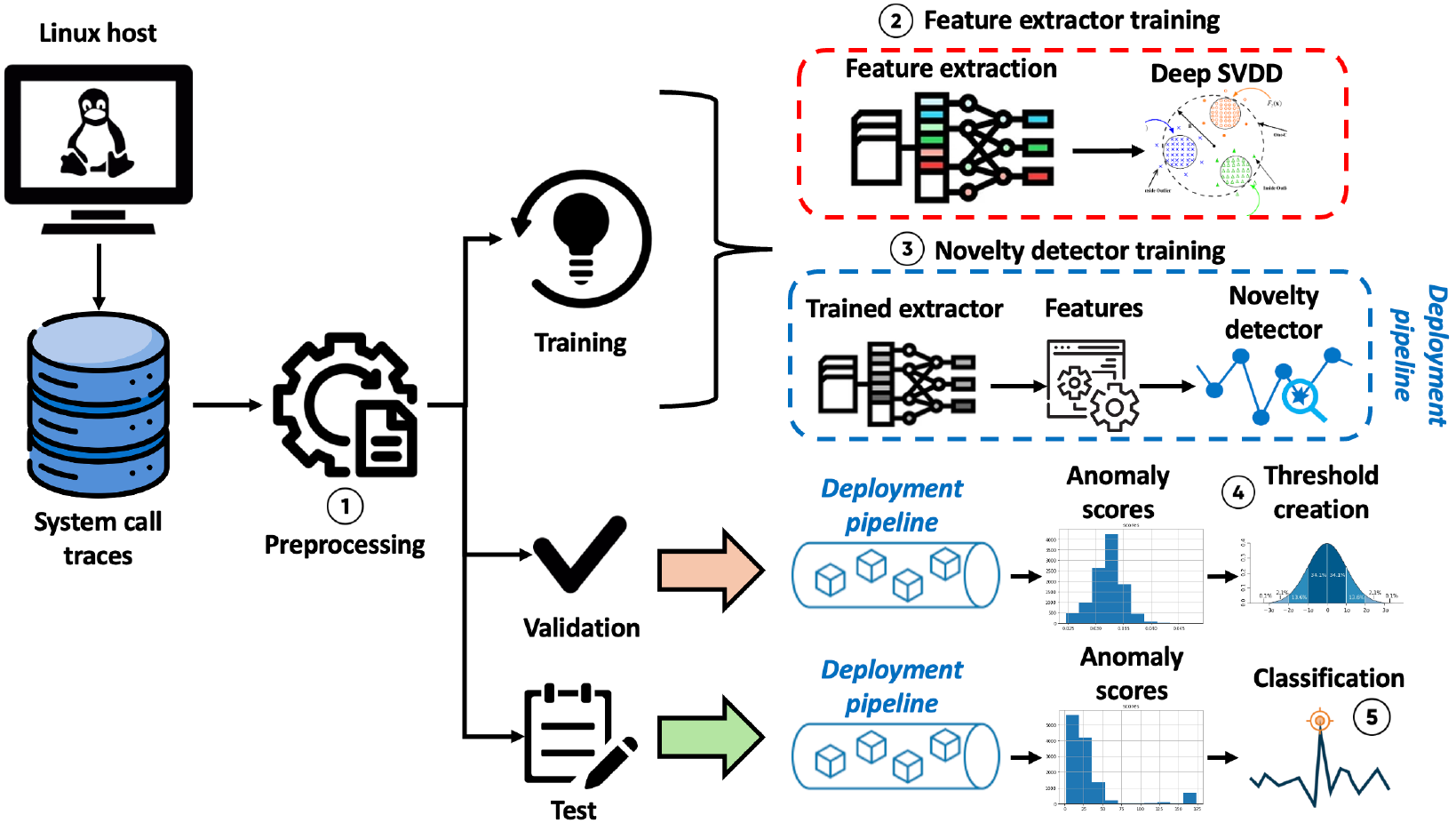}
    \caption{Proposed Lightweight Machine Learning-Based Host Intrusion Detection System (LIGHT-HIDS)}
    \label{fig:framework}
\end{figure*}

We propose LIGHT-HIDS, a lightweight host-based intrusion detection system designed for resource-constrained edge environments. As shown in Figure~\ref{fig:framework}, LIGHT-HIDS adopts a hybrid anomaly detection pipeline that balances efficiency and detection performance. The framework is comprised of  five principal modules: \textcircled{1} data preprocessing, \textcircled{2} feature extractor training, \textcircled{3} novelty detector training, \textcircled{4} threshold creation, and \textcircled{5} anomaly classification. Integer-mapped system call sequences are first processed by a neural network-based feature extractor, and the resulting embeddings are fed into a novelty detector that outputs anomaly scores. These scores are then thresholded to classify sequences as benign or malicious. When applied in real-world scenarios, system call sequences can be tokenized and passed through this pipeline to generate a binary classification indicating the sequence’s safety. Next, we describe the main modules of LIGHT-HIDS in detail.  

\subsection{Data Preprocessing}
Prior to model training, the raw system call sequences were subjected to several preprocessing steps. We first extracted the system call traces, discarding auxiliary metadata such as process IDs and timestamps. Each unique system call name was then tokenized by mapping it to a distinct integer ID, producing a numerical representation suitable for neural network input. Our method follows an anomaly detection paradigm where models are trained solely on normal data to learn typical system behavior. Accordingly, the training and validation sets contain only normal sequences, while the test set includes both normal and anomalous samples to enable comprehensive evaluation of detection performance. 

\subsection{Feature Extractor Training}
To learn robust feature representations from system call sequences, we selected a neural network training framework specifically designed to optimize the feature extractor for anomaly detection. This motivated the adoption of Deep Support Vector Data Description (DeepSVDD), a deep learning-based anomaly detection framework \cite{Ruff2018}. DeepSVDD extends the classical SVDD approach by integrating a neural network. Within this framework, the neural network functions as a feature extractor that maps raw input data, such as sequences of system calls, into a lower-dimensional latent space where representations of normal data are enclosed within a compact hypersphere. The model jointly optimizes the radius of this hypersphere, minimizing its volume to tightly encapsulate the normal data, while simultaneously updating the parameters of the feature extractor through backpropagation. This joint optimization facilitates the learning of feature representations that are intrinsically suited for anomaly detection. 

Our feature extractor architecture begins with a high-dimensional embedding space representing each system call in the input sequence. The feature space is progressively reduced along the sequence dimension using a convolutional neural network (CNN) composed of ReLU-activated one-dimensional convolutional (Conv1D) layers, followed by max pooling. After the sequence dimension is fully reduced, the feature extractor concludes with a fully connected layer. This layer projects the remaining filter outputs into the target feature space, generating the feature vectors that are subsequently input to the DeepSVDD objective. To facilitate deployment in resource-constrained environments, the feature extractor is further optimized using TensorFlow Lite (TFLite)~\cite{tensorflowlite} compression with quantization, significantly reducing model size and improving inference efficiency.

\subsection{Novelty Detector Training}
After training the feature extractor using the DeepSVDD framework, we use it to generate latent representations for the training and validation sequences. These feature vectors are then provided to a lightweight, unsupervised anomaly detection model. Specifically, we adopt Isolation Forest~\cite{liu2008isolation}, which serves as the novelty detector in our framework due to its efficiency and suitability for resource-constrained environments. To train the novelty detector, the original training sequences are first transformed into feature vectors using the trained feature extractor, and these vectors are then used to fit the novelty detection model. Once the novelty detector is trained on the feature vectors, it can assign an anomaly score to each new input vector. This establishes a ``deployment pipeline'', wherein a sequence of integer-mapped system calls is processed through the trained feature extractor to generate a corresponding feature vector, which is then evaluated by the fitted novelty detector to produce an anomaly score. This end-to-end framework enables lightweight and efficient inference from raw system call sequences to anomaly detection outputs.

\subsection{Threshold Creation}
For evaluation, anomaly scores are first obtained by passing the validation sequences, which consist exclusively of normal data, through the proposed ``deployment pipeline''. Based on the distribution of the resulting anomaly scores, we define a fixed threshold to convert the continuous scores into binary anomaly labels, where a label of 0 denotes benign behavior and 1 indicates the presence of an anomaly. This binarization process is essential for downstream evaluation using classification-based metrics such as precision, recall, and F1-score. Specifically, the threshold is chosen as two standard deviations above the mean of the anomaly score distribution, since Isolation Forest assigns higher scores to inputs that are more likely to be anomalous. This method relies on the assumption that the score distribution of normal data is approximately Gaussian and is intended to capture statistical outliers that are indicative of abnormal behavior. 

\subsection{Classification via Threshold}
The predetermined threshold is employed as a decision boundary to classify sequences within the unseen test set. Each test sequence is processed through the deployment pipeline to yield an anomaly score. Sequences exhibiting scores above the established threshold are classified as anomalous, whereas those with scores below the threshold are classified as benign. This thresholding mechanism provides a standardized procedure for converting continuous anomaly scores into binary classifications. Subsequently, the predicted labels are compared against ground truth annotations to rigorously assess the detection performance of the model.

%% file: sections/Experiment.tex
\subsection{Experimental Setup}
\subsubsection{Hardware}
Model training was conducted on a single NVIDIA GTX 2080 Ti GPU. For inference, we used CPU-only execution on an Intel Xeon Gold 6230 to simulate the lack of GPU acceleration typically encountered in resource-constrained edge devices. Additionally, we evaluate our models on the NVIDIA Jetson Orin NX with 16 GB of RAM~\cite{nvidia2024orin}, a representative platform for real-world edge deployments.  



\subsubsection{Datasets} 
We evaluate our approach using two subsets from the Leipzig Intrusion Detection Data Set (LID-DS)~\cite{Grimmer2023}, a modern, system call-based dataset designed for anomaly-based HIDS research. The first subset corresponds to brute-force attacks, which constitute approximately 22\% of attacks on edge gateways~\cite{Xiao2019}. These attacks involve repeated, systematic attempts to guess login credentials, generating distinctive system call patterns due to frequent authentication failures. The second subset focuses on SQL injection attacks, a representative class of malware injection attacks that account for 26\% of edge threats~\cite{Grimmer2023}. These attacks manipulate input to inject malicious SQL queries into applications, often causing harmful database operations and system behavior that can be captured at the system call level. Both subsets include labeled sequences of benign and malicious activity, facilitating comprehensive evaluation in an anomaly detection context. Together, these attack types cover a substantial portion of potential host-based threats, offering strong potential for generalizing HIDS performance to other attack categories.

\subsubsection{Evaluation Metrics}
Our evaluation focuses on two key aspects: detection performance and inference time.

\begin{itemize}
\item \textbf{Detection Performance:} Precision, recall, and F1-score to assess the model’s ability to correctly identify malicious activity while minimizing false alarms.
\item \textbf{Inference Time:} The elapsed time from inputting system call sequences to generating anomaly predictions, reflecting the model’s computational overhead.
\end{itemize}

The primary goal is to show that our method maintains high detection accuracy while significantly lowering inference latency, enabling timely intrusion detection.

\subsubsection{Baselines}
We evaluate our approach against three categories of baselines: host-based intrusion detection (HIDS) methods, time-series anomaly detection models, and classical anomaly detection techniques.

\begin{itemize}
    \item \textbf{HIDS methods:} We implemented several deep learning models commonly used in host-based intrusion detection, including WaveNet, LSTM, and CNNRNN hybrid, following the methodology of Ring et al.~\cite{Ring2021}. These models use language modeling techniques to predict the next system call in a sequence. For training, input-target pairs were constructed by shifting system call sequences by one position. During inference, anomaly scores were derived from the model's prediction certainty, calculated using the maximum softmax probabilities across the sequence.
    \item \textbf{Time-series anomaly detection:} We compare our approach against leading time-series anomaly detection models, including COUTA~\cite{xu2024calibrated}, TimesNet~\cite{wu2022timesnet}, and Deep SVDD~\cite{Ruff2018}. These baselines were implemented using the DeepOD library~\cite{xu2023deep,xu2024calibrated}.
    \item \textbf{Classical anomaly detection:} We also evaluate traditional models, including Isolation Forest (IF)~\cite{liu2008isolation} and One-Class SVM (OCSVM)~\cite{bounsiar2014one}, which output anomaly scores analogous to those produced by our language modeling baselines. For input, we use average-pooled feature embeddings extracted from the LSTM model proposed in~\cite{Ring2021}, resulting in fixed-length vectors matching the dimensionality of the system call embedding space.
\end{itemize}


\begin{figure}[]
    \centering
    \begin{subfigure}[b]{0.97\columnwidth}
        \includegraphics[width=\linewidth]{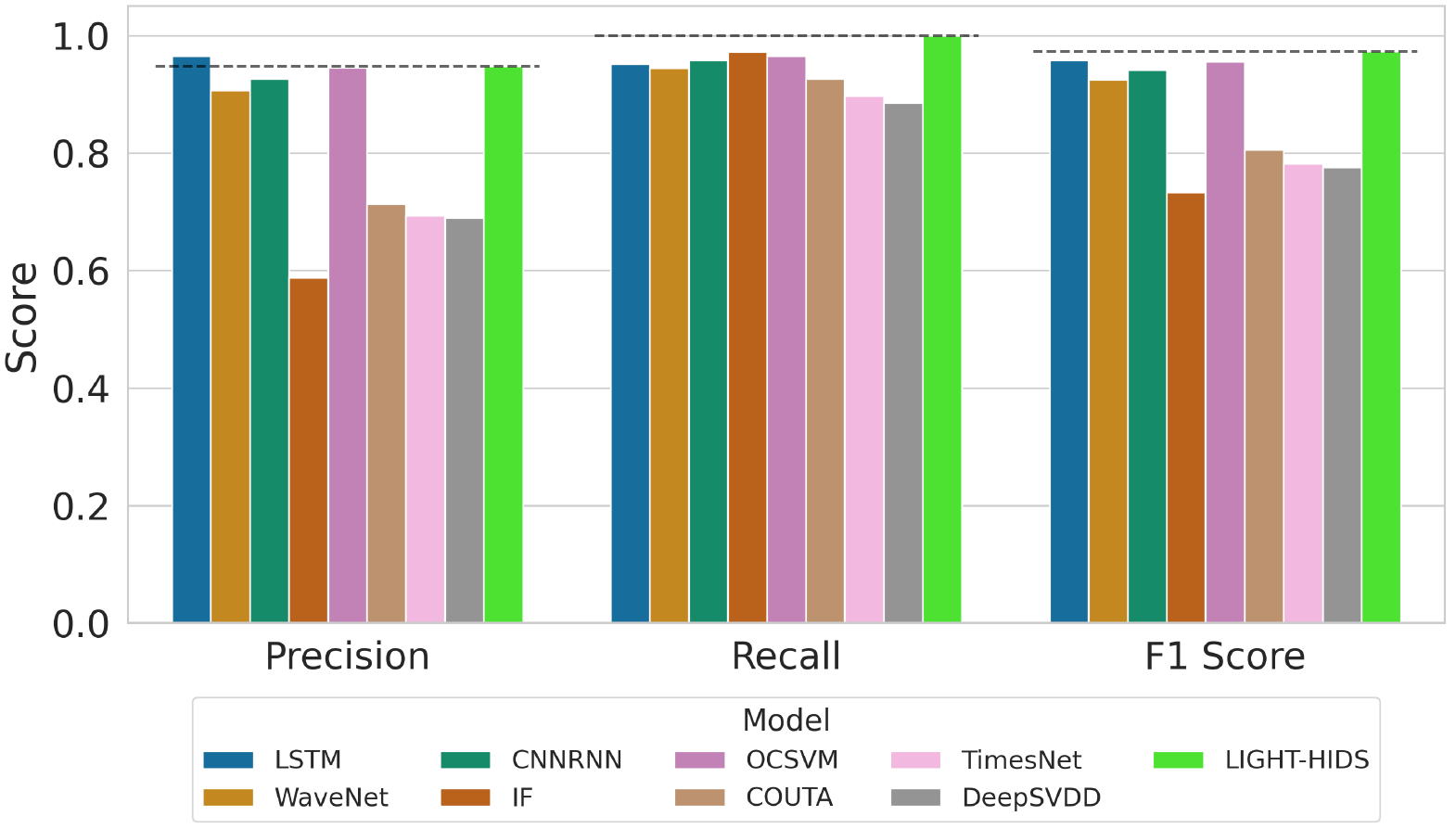}
        \caption{Brute-force attack}
        \label{fig:bruteforce_results}
    \end{subfigure}
    \hfill
    \begin{subfigure}[b]{0.97\columnwidth}
        \includegraphics[width=\linewidth]{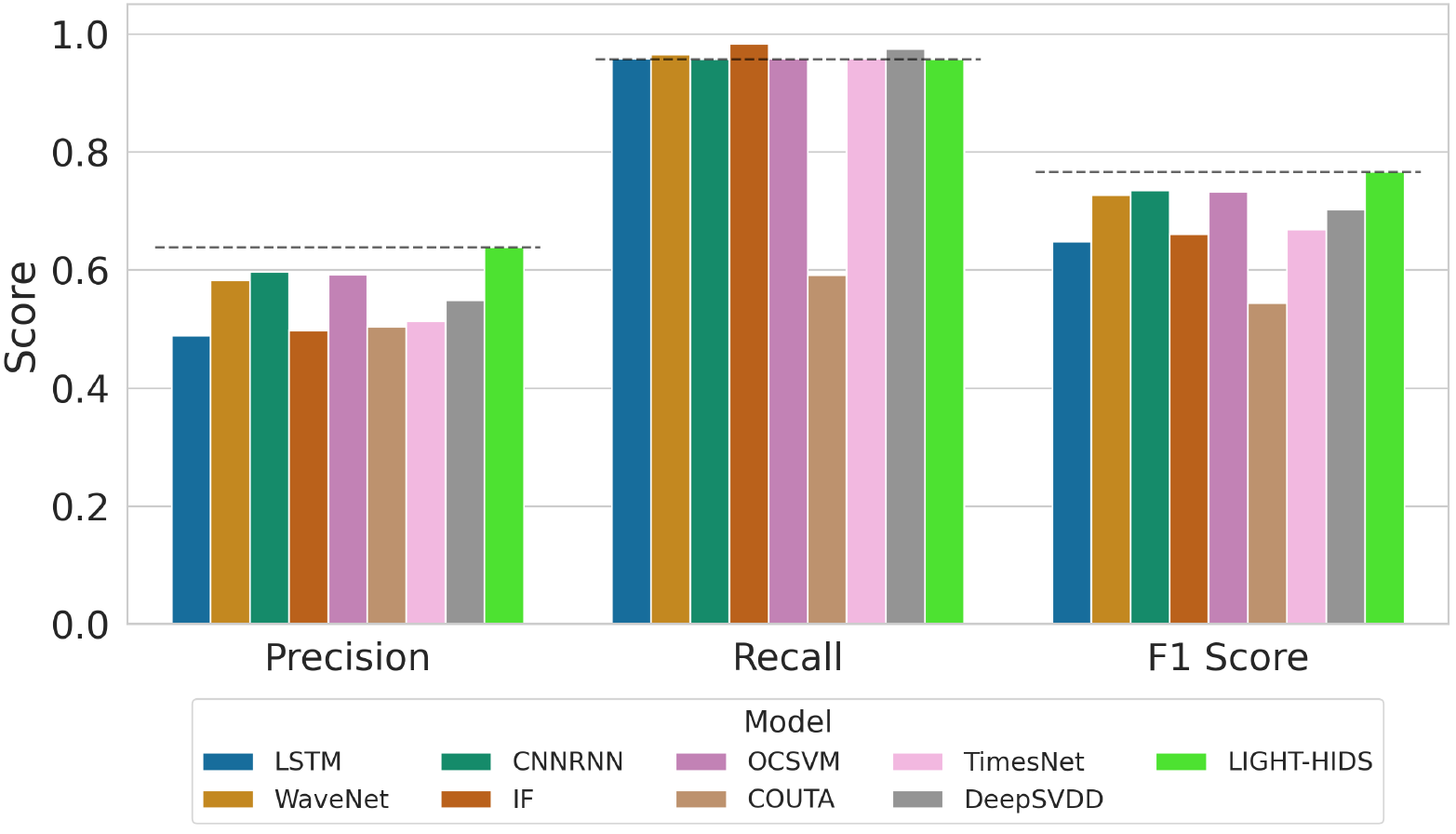}
        \caption{SQL injection attack}
        \label{fig:sql_results}
    \end{subfigure}
    \caption{Performance Comparison with State-of-the-Art}
    \label{fig:results}
\end{figure}

\subsection{Results}
\subsubsection{Detection Performance}
Figure~\ref{fig:results} presents a comparative analysis of the anomaly detection performance of our proposed method against state-of-the-art models on the brute-force (Figure~\ref{fig:bruteforce_results}) and SQL injection (Figure~\ref{fig:sql_results}) datasets. In each subfigure, the x-axis represents the evaluation metrics (precision, recall, and F1 score) while the y-axis indicates the corresponding values. Different colors correspond to the various baseline methods, with LIGHT-HIDS highlighted in light green as the rightmost column. We observe that \textbf{LIGHT-HIDS achieves the highest F1 score across both datasets}, indicating robust performance in distinguishing intrusive activity from normal behavior relative to the baseline models. Although LIGHT-HIDS does not consistently outperform all baselines on every individual metric, such as recall on the SQL dataset, it maintains competitive performance by remaining within 0.05 of the top-performing models in these cases. Furthermore, methods specifically designed for HIDS exhibit better performance relative to other baseline categories. 

\begin{table}[]
    \centering
    \caption{F1 Score Comparison by Baseline Category}
    \label{tab:f1_score_all_categories}
    \begin{tabular}{lcccc}
        \toprule
        \textbf{Dataset} & \textbf{Category} & \textbf{Best Score} & \textbf{LIGHT-HIDS} & \textbf{$\uparrow$ (\%)} \\
        \midrule
        \multirow{3}{*}{Brute Force} 
            & HIDS       & 0.958 &  & 1.57 \\
            & Time-Series              & 0.775 & \textbf{0.973} & \textbf{25.48} \\
            & Traditional & 0.955 & & 1.85 \\
        
        \midrule
        \multirow{3}{*}{SQL Injection} 
            & HIDS     & 0.735 &  & 4.23 \\
            & Time-Series             & 0.702 & \textbf{0.766} & \textbf{9.03} \\
            & Traditional & 0.732 &  & 4.59 \\
        \bottomrule
    \end{tabular}
\end{table}

To quantify improvements in F1 score, Table~\ref{tab:f1_score_all_categories} reports LIGHT-HIDS’s gains over the best-performing baseline within each category. For the brute-force dataset, the top models are LSTM (HIDS baselines), COUTA (time-series anomaly detection), and OCSVM (traditional anomaly detection), while for the SQL injection dataset, the leading models are CNNRNN, DeepSVDD, and OCSVM, respectively. LIGHT-HIDS demonstrates notable improvements on the brute-force dataset, ranging from 1.57\% to 25.48\%, with the largest gain observed over the time-series baseline. Similarly, on the SQL injection dataset, improvements range from 4.23\% to 9.03\%, again highlighting superiority over time-series methods. Table~\ref{tab:f1_score_best_baselines} further compares LIGHT-HIDS to the top three performing models per dataset, showing that HIDS-specific methods and OCSVM represent the strongest baselines.

When all baselines are taken into consideration, LIGHT-HIDS achieves up to \textbf{32.74\%} improvement with an average gain of \textbf{16.24\%} on the brute-force dataset, and up to \textbf{40.83\%} improvement with an average gain of \textbf{14.11\%} on the SQL injection dataset. These consistent gains across attack types and model categories underscore the robustness and effectiveness of LIGHT-HIDS in enhancing host-based intrusion detection.


\begin{table}[]
    \centering
    \caption{F1 Score Comparison over Best Baselines}
    \label{tab:f1_score_best_baselines}
    \begin{tabular}{lcccc}
        \toprule
        \textbf{Dataset} & \textbf{Model} & \textbf{F1 Score} & \textbf{LIGHT-HIDS} & \textbf{$\uparrow$ (\%)} \\
        \midrule
        \multirow{3}{*}{Brute Force} 
            & LSTM        & 0.958 &  & 1.57 \\
            & OCSVM              & 0.955 & \textbf{0.973} & 1.85 \\
            & CNNRNN  & 0.942 & & 3.29 \\
        
        \midrule
        \multirow{3}{*}{SQL Injection} 
            & CNNRNN     & 0.735 &  & 4.23 \\
            & OCSVM             & 0.732 & \textbf{0.766} & 4.60 \\
            & WaveNet & 0.726 &  & 5.42 \\
        \bottomrule
    \end{tabular}
\end{table}

\subsubsection{Inference Time}
Table~\ref{tab:light_hids_speedup_combined} presents the inference-time speedup of LIGHT-HIDS relative to the best-performing baseline models on each dataset, as identified in Table~\ref{tab:f1_score_best_baselines}. Evaluations were conducted on both an Intel Xeon CPU and an NVIDIA Jetson Orin NX device. On the Brute Force dataset, LIGHT-HIDS achieves up to 33.77$\times$ faster inference than CNNRNN on the CPU, while maintaining a superior detection accuracy (F1 score of 0.973 compared to 0.942). Similarly, for SQL Injection, LIGHT-HIDS delivers 75$\times$ faster inference than CNNRNN, alongside a 4.23\% improvement in F1 score. Comparable trends are observed on the Jetson, where LIGHT-HIDS significantly reduces inference latency relative to baseline models while sustaining competitive accuracy. While models such as LSTM and OCSVM demonstrate competitive F1 scores on Brute Force (0.958 and 0.955, respectively), LIGHT-HIDS achieves higher accuracy and 3.6 to 6.4$\times$ faster inference on the CPU. For SQL Injection, LIGHT-HIDS outperforms OCSVM and WaveNet by 4.6\% and 5.42\% in F1, while offering nearly 10$\times$ faster inference. 

The Intel Xeon CPU achieves faster inference due to its higher processing power, whereas the NVIDIA Jetson Orin NX balances inference speed with energy efficiency, highlighting important trade-offs in deploying host intrusion detection systems on edge hardware. \textbf{These results show that LIGHT-HIDS delivers high detection accuracy with substantially reduced inference latency, making it well-suited for real-time host intrusion detection across both high-performance servers and resource-constrained edge devices.}

\begin{table*}[t]
\centering
\caption{Inference Time and Speedup of LIGHT-HIDS on Intel Xeon Gold 6230 (CPU-only) and NVIDIA Jetson Orin NX}
\label{tab:light_hids_speedup_combined}
\begin{tabular}{lcccccccc}
\toprule
\multirow{2}{*}{\textbf{Model}} & \multicolumn{4}{c}{\textbf{Brute Force}} & \multicolumn{4}{c}{\textbf{SQL Injection}} \\
\cmidrule(lr){2-5} \cmidrule(lr){6-9}
 & \multicolumn{2}{c}{\textbf{Xeon (CPU)}} & \multicolumn{2}{c}{\textbf{Jetson}} & \multicolumn{2}{c}{\textbf{Xeon (CPU)}} & \multicolumn{2}{c}{\textbf{Jetson}} \\
 & Time (s) & Speedup & Time (s) & Speedup & Time (s) & Speedup & Time (s) & Speedup \\
\midrule
LIGHT-HIDS & 0.004 & -- & 0.030 & -- & 0.030 & -- & 0.293 & -- \\
LSTM       & 0.016 & 3.61$\times$ & 0.420 & 13.90$\times$ & 0.113 & 3.76$\times$ & 4.310 & 14.70$\times$ \\
WaveNet    & 0.073 & 16.57$\times$ & 0.116 & 3.84$\times$ & 0.295 & 9.83$\times$ & 0.616 & 2.10$\times$ \\
CNNRNN     & 0.149 & \textbf{33.77$\times$} & 1.011 & \textbf{33.47$\times$} & 2.248 & \textbf{74.93$\times$} & 10.939 & \textbf{37.31$\times$} \\
OCSVM      & 0.028 & 6.41$\times$ & 0.056 & 1.87$\times$ & 0.288 & 9.60$\times$ & 0.494 & 1.68$\times$ \\
\bottomrule
\end{tabular}
\end{table*}

%% file: sections/Conclusion.tex
The rapid expansion of IoT and edge computing has introduced significant security challenges alongside new opportunities. Machine learning (ML)–based Host Intrusion Detection Systems (HIDS) are critical for accurately detecting complex attack patterns. However, most state-of-the-art ML-based solutions rely on computationally heavy models that hinder practical deployment in resource-constrained edge environments. This creates a pressing need for lightweight, efficient ML approaches that maintain high detection accuracy. To address these limitations, we present LIGHT-HIDS, an efficient host-based intrusion detection framework that employs a compressed neural network to extract compact and meaningful features from system call data. This feature extractor is paired with a lightweight novelty detection model, enabling accurate and fast anomaly detection. Our results show that LIGHT-HIDS improves detection accuracy while reducing inference time by up to 75$\times$ compared to state-of-the-art methods, demonstrating its potential as a scalable solution for real-time intrusion detection in resource-constrained environments.